\documentclass
[floatfix,superscriptaddress,secnumarabic,amssymb,amsmath,nobibnotes,aps,prd,showkeys,showpacs,nofootinbib,onecolumn,notitlepage,12pt]{revtex4-2}%
\usepackage{setspace}
\usepackage{color}
\usepackage{amsmath}
\usepackage{amsfonts}
\usepackage{verbatim}
\usepackage{amssymb}
\usepackage{graphicx,bm}
\usepackage{graphicx}
\usepackage{amsmath}
\usepackage{amssymb}
\usepackage{amssymb}
\usepackage{graphicx,bm}
\usepackage{graphicx}
\usepackage[caption=false]{subfig}%
\providecommand{\U}[1]{\protect\rule{.1in}{.1in}}

\newcommand{\be}{\begin{equation}}
\newcommand{\ee}{\end{equation}}

\newcommand{\mincir}{\raise
-3.truept\hbox{\rlap{\hbox{$\sim$}}\raise4.truept\hbox{$<$}\ }}
\newcommand{\magcir}{\raise
-3.truept\hbox{\rlap{\hbox{$\sim$}}\raise4.truept\hbox{$>$}\ }}

\ifx\pdfoutput\relax\let\pdfoutput=\undefined\fi
\newcount\msipdfoutput
\ifx\pdfoutput\undefined\else
\ifcase\pdfoutput\else
\ifx\paperwidth\undefined\else
\ifdim\paperheight=0pt\relax\else\pdfpageheight\paperheight\fi
\ifdim\paperwidth=0pt\relax\else\pdfpagewidth\paperwidth\fi
\fi\fi\fi
\begin{document}
\title{Minisuperspace description of $f(Q)$-cosmology}
\author{A. Paliathanasis}
\email{anpaliat@phys.uoa.gr}
\affiliation{Institute of Systems Science, Durban University of Technology, Durban 4000,
South Africa}
\affiliation{Departamento de Matem\'{a}ticas, Universidad Cat\'{o}lica del Norte, Avda.
Angamos 0610, Casilla 1280 Antofagasta, Chile}
\author{N. Dimakis}
\email{nikolaos.dimakis@ufrontera.cl}
\affiliation{Departamento de Ciencias F\'isicas, Universidad de la Frontera, Casilla 54-D,
4811186 Temuco, Chile}
\author{T. Christodoulakis}
\email{tchris@phys.uoa.gr}
\affiliation{Nuclear and Particle Physics section, Physics Department, University of
Athens, 15771 Athens, Greece}

\begin{abstract}
We investigate the existence of minisuperspace description for the homogeneous
cosmological field equations within the framework of symmetric teleparallel
$f(Q)$-gravity. We consider the background space to be described by the
isotropic Friedmann--Lema\^{\i}tre--Robertson--Walker geometry, the
anisotropic Kantowski-Sachs and the anisotropic Bianchi III geometries. Across
all these models, we establish that the field equations in $f(Q)$-cosmology
exhibit second-order characteristics in the coincident gauge and those of a
sixth-order theory in the non-coincident gauge. Specifically, within the
latter scenario, the dynamic degrees of freedom are attributed to two scalar
fields. Finally, as an example of integrability, we derive a vacuum
cosmological solution within the non-coincident gauge.

\end{abstract}
\maketitle

\section{Introduction}

\label{sec1}

The concept of a minisuperspace description is essential within
gravitational theories, as it allows us a better understanding of the dynamics and permits a
physical description on geometrodynamical terms - when physical observables, for example, are associated with symmetries of the minisuperspace.
Moreover, when a minisuperspace exists, the gravitational field equations can be expressed in a
more simplified manner \cite{mp1}, so as to give to certain degrees of freedom an equivalent scalar field
description. In addition to the above, the establishment of a point-like Lagrangian
function, facilitates the utilization of mathematical
tools and techniques from analytical mechanics, enabling the derivation and
construction of exact and analytic solutions for the field equations
\cite{ns1,ns2,ns3,ns4,ns5,ns6}. The equivalent Hamiltonian formulation, through the use of the Dirac-Bergmann algorithm \cite{DB1,DB2}, permits the distinction of the true degrees of freedom for the reduced gravitational system. This is quite useful, especially in cases where the generic Hamiltonian formalism for the full gravitational theory, is not so straightforwardly derivable; as happens in the case of $f(Q)$ gravity, see the relative discussions over the physical degrees of freedom \cite{hd1,hd2,hd3}.

Apart from the tools and simplifications that a minisuperspace description offers at the classical regime, it is also of great importance in the
realm of quantum cosmology \cite{ns7}. The latter is our best alternative to study quantum aspects of gravitational systems, due to the lack of a complete theory of quantum gravity. In two separate studies, DeWitt
\cite{deWitt} and Wheeler \cite{whe} introduced a scheme for a canonical quantization of gravity, leading to the creation of the so-called Wheeler-DeWitt equation.
Within the $3+1$ decomposition of General Relativity, the
Wheeler-DeWitt equation is derived from the Hamiltonian constraint of the
theory. In its full generality, is actually a three-fold infinite family of
equations on 3-dimensional hypersurfaces. Yet, when a minisuperspace
description exists, the Wheeler-DeWitt equation is streamlined into a single
equation, analogous to the Schr\"{o}dinger equation of quantum mechanics
\cite{ha1,ha2}.

In this study, we investigate the minisuperspace description of the field
equations within $f(Q)$-gravity \cite{fq}, focusing on a homogeneous and
isotropic Friedmann--Lema\^{\i}tre--Robertson--Walker geometry (FLRW).
$f(Q)$-gravity represents an extension of Symmetric Teleparallel General
Relativity (STGR) \cite{ngr2}, where the gravitational Lagrangian takes the
form of a nonlinear function involving the nonmetricity scalar derived from a
symmetric and flat connection.

While General Relativity (GR) is defined by the metric tensor and the
Levi-Civita connection giving rise to the Ricci scalar $R$, Symmetric
Teleparallel General Relativity (STGR) adopts a different approach. In STGR,
the essential geometric objects are the metric tensor and the symmetric flat
connection that gives rise to the non-metricity scalar $Q$. On the other hand,
in teleparallelism \cite{ngr1}, the foundational geometric quantities consist
of the vierbein fields and the Weitzenb\"{o}ck connection, which in turn
yields the torsion scalar $T$ \cite{Weitzenb23}. Several works exist studying
this geometric trinity of gravitational theories and their modifications
\cite{Lav00,Lav0a,Lav0b,Lav1,Lav2,VDF1,VDF2,Lav3,Koivst0,Koivst}.

Recently, $f(Q)$-cosmology has attracted considerable attention due to its
ability to offer a straightforward geometric mechanism for explaining the
late-time acceleration of the universe. This framework also challenges the
conventional $\Lambda$-cosmology when it comes to analyzing cosmological data
\cite{ds00,ww0,ww8}. A novel approach to study quantum cosmology in the
framework of the coincident gauge in $f\left(  Q\right)  $-theory is presented
in \cite{dm1}. The works \cite{dm2,dm3} put forth exact solutions in
$f(Q)$-cosmology for both isotropic and anisotropic homogeneous spacetimes.
The behavior of cosmological parameters in asymptotic regimes is outlined in
\cite{dm4,dm5,dm6,dm7}. An interesting discovery in the context of the
coincident gauge is that inflation and late-time acceleration are inherently
provided by the theory, negating the necessity for introducing a cosmological
constant term or exotic matter sources \cite{dm7}. This can be attributed to
the presence of dynamic degrees of freedom in the field equations associated
with boundary terms arising from the connection.

Furthermore, the analysis of cosmological parameter trajectories in the
presence of spatial curvature indicates that $f(Q)$-theory has the potential
to address the flatness problem \cite{dm8}. In \cite{sl1,sl2}, cosmological
models were introduced within $f(Q)$-theory, capable of describing phantom
epochs. For a broader range of studies in $f(Q)$-cosmology, interested readers
can refer to \cite{sl3,sl4,sl5,sl6} and the references therein. Efforts to
discuss the Hamiltonian formalism of $f(Q)$-gravity and the ADM formalism are
presented in \cite{hd1,hd2,hd3}. For a review in $f\left(  Q\right)  $-theory
we refer the reader in \cite{revh}. The organization of the paper is outlined
in the following lines.

The basic properties and definitions of $f\left(  Q\right)  $-gravity are
outlined in Section \ref{sec2}. In Section \ref{sec3} we consider an isotopic
and homogeneous geometry, presenting a comprehensive collection of the field
equation sets derived from various families of connections. For each of these
four distinct connection families, we offer the corresponding point-like
Lagrangian functions. Our analysis leads to the inference that opting for a
connection defined within the non-coincident gauge, $f(Q)$-cosmology is
characterized by the presence of two scalar fields. The\ field equations in
the scalar field equivalent description are given. Moreover, in Section
\ref{sec4} we take advantage of the scalar field representation for the field
equations. In the context of spatially flat geometry, we streamline the field
equations into a singular second-order differential equation. We also present
the analytic solution for the power-law model $f\left(  Q\right)  =Q^{\alpha}%
$. In Section \ref{sec5} we extend our analysis and we search for a
minisuperspace description for anisotropic background spaces. In particular we
consider the Kantowski-Sachs spacetime and the Bianchi III geometry and it
follows that for the frame defined in non-coincident gauge a minisuperspace
description exist in $f\left(  Q\right)  $-theory which extends the
minisuperspace for the FLRW geometry with non-zero spatial curvature, In
Appendix \ref{appen} we show how a Lagrange multiplier can be introduced in
order to reconstruct the point-like Lagrangian functions. Finally, our
findings are summarized in Section \ref{conc}.

\section{$f\left(  Q\right)  $-gravity}

\label{sec2}

Consider a four-dimensional manifold characterized by the metric tensor
$g_{\mu\nu}$ and the generic connection $\Gamma_{\mu\nu}^{\kappa}$ which
defines the covariant derivative $\nabla_{\lambda}$. Consequently, using this
connection, we can formulate the curvature tensor $R_{\;\lambda\mu\nu}%
^{\kappa}$, the torsion $T_{\mu\nu}^{\lambda}$ tensor and the non-metricity
$Q_{\lambda\mu\nu}$ tensor as follows \cite{Eisenhart}
\begin{align}
R_{\;\lambda\mu\nu}^{\kappa}  &  =\frac{\partial\Gamma_{\;\lambda\nu}^{\kappa
}}{\partial x^{\mu}}-\frac{\partial\Gamma_{\;\lambda\mu}^{\kappa}}{\partial
x^{\nu}}+\Gamma_{\;\lambda\nu}^{\sigma}\Gamma_{\;\mu\sigma}^{\kappa}%
-\Gamma_{\;\lambda\mu}^{\sigma}\Gamma_{\;\mu\sigma}^{\kappa}\\
\mathrm{T}_{\mu\nu}^{\lambda}  &  =\Gamma_{\;\mu\nu}^{\lambda}-\Gamma
_{\;\nu\mu}^{\lambda}\\
Q_{\lambda\mu\nu}  &  =\nabla_{\lambda}g_{\mu\nu}=\frac{\partial g_{\mu\nu}%
}{\partial x^{\lambda}}-\Gamma_{\;\lambda\mu}^{\sigma}g_{\sigma\nu}%
-\Gamma_{\;\lambda\nu}^{\sigma}g_{\mu\sigma}.
\end{align}

In the STGR, connection $\Gamma_{\mu\nu}^{\kappa}$ has the property of being
flat, i.e. $R_{\;\lambda\mu\nu}^{\kappa}=0$ and torsionless, $\mathrm{T}%
_{\mu\nu}^{\lambda}=0$.

Then, from the non-metricity tensor $Q_{\lambda\mu\nu}$ we can construct the
scalar \cite{ngr2}
\[
Q=Q_{\lambda\mu\nu}P^{\lambda\mu\nu}%
\]
where \cite{ngr2}%
\begin{equation}
P_{\;\mu\nu}^{\lambda}=-\frac{1}{4}Q_{\;\mu\nu}^{\lambda}+\frac{1}{2}%
Q_{(\mu\phantom{\lambda}\nu)}^{\phantom{(\mu}\lambda\phantom{\nu)}}+\frac
{1}{4}\left(  Q^{\lambda}-\bar{Q}^{\lambda}\right)  g_{\mu\nu}-\frac{1}%
{4}\delta_{\;(\mu}^{\lambda}Q_{\nu)}, \label{defP}%
\end{equation}
which is written with the help of the traces $Q_{\mu}=Q_{\mu\nu}%
^{\phantom{\mu\nu}\nu}$ and $\bar{Q}_{\mu}=Q_{\phantom{\nu}\mu\nu}%
^{\nu\phantom{\mu}\phantom{\mu}}$. Parenthesis in the indices denote
symmetrization, that is, $A_{(\mu\nu)}=\frac{1}{2}\left(  A_{\mu\nu}+A_{\nu
\mu}\right)  $; and $\delta_{\;\nu}^{\mu}$ is the Kroncker delta.

The non-metricity scalar $Q$ and the corresponding Ricci scalar $\overset
{o}{R}$, defined by the Levi-Civita connection for the metric tensor, differ
by a boundary term, which means that%
\begin{equation}
\int d^{4}x\sqrt{-g}Q\simeq\int d^{4}x\sqrt{-g}\overset{o}{R}+\text{boundary
terms.}%
\end{equation}
As a result, STGR is dynamically equivalent to GR.

This scenario changes in the context of nonlinear $f(Q)$-gravity, where the
gravitational Action Integral is defined as \cite{Hohmann,Heis2}
\begin{equation}
S_{f\left(  Q\right)  }=\int d^{4}x\sqrt{-g}f(Q),
\end{equation}
characterized by a function $f(Q)$ which is no longer linear in $Q$.

Varying the Action Integral $S_{f\left(  Q\right)  }$ with respect to the
metric tensor yields the modified field equations \cite{Hohmann,Heis2}
\begin{equation}
\frac{2}{\sqrt{-g}}\nabla_{\lambda}\left(  \sqrt{-g}f^{\prime}(Q) P_{\;\mu\nu
}^{\lambda}\right)  -\frac{1}{2}f(Q)g_{\mu\nu}+f^{\prime}(Q)\left(  P_{\mu
\rho\sigma}Q_{\nu}^{\;\rho\sigma}-2Q_{\rho\sigma\mu}P_{\phantom{\rho\sigma}\nu
}^{\rho\sigma}\right)  =0, \label{fl1}%
\end{equation}
where, throughout this work, we use the prime to denote differentiation with
respect to the argument of some function; in the above relation $f^{\prime
}(Q)=\frac{df}{dQ}$.

Furthermore, variation of the Action Integral $S_{f\left(  Q\right)  }$ with
respect the symmetric flat connection leads to the equation of motion for the
connection \cite{Hohmann,Heis2}
\begin{equation}
\nabla_{\mu}\nabla_{\nu}\left(  \sqrt{-g}f^{\prime}(Q)
P_{\phantom{\mu\nu}\sigma}^{\mu\nu}\right)  =0. \label{kl.01}%
\end{equation}

If the latter equation is fulfilled identically, we will designate this
connection as being in the ``coincident gauge''. Although formally, the
coincident gauge corresponds to when the components of the connection are
zero, here we shall use the term in a more relaxed sense, signifying a
connection which is nondynamical, i.e. a connection for which \eqref{kl.01} is
a trivial identity. In the ``non-coincident gauge'', the differential equation
(\ref{kl.01}), as we are going to see, will introduce new dynamic degrees of freedom.

\section{FLRW Cosmology}

\label{sec3}

On large scales, the universe exhibits isotropic and homogeneous
characteristics, described by the FLRW line element. In spherical coordinates,
this line element reads:
\begin{equation}
ds^{2}=-N(t)^{2}dt^{2}+a(t)^{2}\left[  \frac{dr^{2}}{1-kr^{2}}+r^{2}\left(
d\theta^{2}+\sin^{2}\theta d\varphi^{2}\right)  \right]  , \label{genlineel}%
\end{equation}
where $a\left(  t\right)  $ is the scale factor and $N\left(  t\right)  $ is
the lapse function. The Hubble function is defined as $H=\frac{1}{N}\frac
{\dot{a}}{a}$, where $\dot{a}=\frac{da}{dt}$. Spacetime (\ref{genlineel})
admits six symmetries comprised by the three vectors
\begin{equation}
\zeta_{1}=\sin\varphi\partial_{\theta}+\frac{\cos\varphi}{\tan\theta}%
\partial_{\varphi},\quad\zeta_{2}=-\cos\varphi\partial_{\theta}+\frac
{\sin\varphi}{\tan\theta}\partial_{\varphi},\quad\zeta_{3}=-\partial_{\varphi}
\label{Kil1}%
\end{equation}
and the three rotations
\begin{equation}%
\begin{split}
\xi_{1}  &  =\sqrt{1-kr^{2}}\sin\theta\cos\varphi\partial_{r}+\frac
{\sqrt{1-kr^{2}}}{r}\cos\theta\cos\varphi\partial_{\theta}-\frac
{\sqrt{1-kr^{2}}}{r}\frac{\sin\varphi}{\sin\theta}\partial_{\varphi}\\
\xi_{2}  &  =\sqrt{1-kr^{2}}\sin\theta\sin\varphi\partial_{r}+\frac
{\sqrt{1-kr^{2}}}{r}\cos\theta\sin\varphi\partial_{\theta}+\frac
{\sqrt{1-kr^{2}}}{r}\frac{\cos\varphi}{\sin\theta}\partial_{\varphi}\\
\xi_{3}  &  =\sqrt{1-kr^{2}}\cos\theta\partial_{r}-\frac{\sqrt{1-kr^{2}}}%
{r}\sin\theta\partial_{\varphi}.
\end{split}
\label{Kil2}%
\end{equation}

The general structure of all compatible symmetric connections,
possessing the symmetries of the underlying spacetime and resulting in a
curvature tensor of zero magnitude, has been previously established in
\cite{Hohmann,Heis2,Zhao}. Specifically, it was revealed that within the
context of the spatially flat FLRW geometry, three connections emerge: one
defined within the coincident gauge, and two within the non-coincident gauge.
However, in the case of $k\neq0$, a solitary connection exist,
defined in the non-coincident gauge. 

Although in symmetric teleparallel theory the selection of the
connection does not affect the gravitational theory, that it is not true for
the case of $f\left(  Q\right)  $-gravity. In the framework of FLRW
geometry, in \cite{dm2}, the effects of using different connections over the
existence of cosmological solutions was studied. It was established, that the selection of the
connection affects the ensuing cosmological history \cite{dm7}. Accelerated solutions, which can describe the
early-time and the late-time acceleration phases, were explicitly derived. In \cite{ds00},
cosmological observations have been used to test the different connections in
$f\left(  Q\right)  $-gravity as dark energy candidates. However, from
these results, it was not possible to select a preferred connection. 

In the following, we outline the field equations corresponding to each
connection, and we also introduce the corresponding point-like Lagrangians
that give rise to the field equations within the framework of the
minisuperspace approach.

\subsection{Spatially flat case $k=0$}

In the context of the spatially flat FLRW geometry the common non-zero
components of the three admissible connections are%
\[
\Gamma_{\theta\theta}^{r}=-r~,~\Gamma_{\varphi\varphi}^{r}=-r\sin^{2}\theta
\]%
\[
\Gamma_{\varphi\varphi}^{\theta}=-\sin\theta\cos\theta~,~\Gamma_{\theta
\varphi}^{\varphi}=\Gamma_{\varphi\theta}^{\varphi}=\cot\theta
\]%
\[
\Gamma_{\;r\theta}^{\theta}=\Gamma_{\;\theta r}^{\theta}=\Gamma_{\;r\varphi
}^{\varphi}=\Gamma_{\;\varphi r}^{\varphi}=\frac{1}{r}%
\]
Here, we can observe that these components correspond to the Levi-Civita
components for the three-dimensional flat space, characterized by the line
element \cite{Zhao}:%
\[
ds_{\left(  3\right)  }^{2}=dr^{2}+r^{2}\left(  d\theta^{2}+\sin^{2}\theta
d\varphi^{2}\right)  .
\]

\subsubsection{Connection $\Gamma_{1}$}

Connection $\Gamma_{1}$ is established within the coincident gauge, featuring
just a single additional non-zero component
\begin{equation}
\Gamma_{\;tt}^{t}=\gamma(t), \label{con1}%
\end{equation}
where $\gamma(t)$ is a function of the time variable $t$.

For $\Gamma_{1}$ we derive, from the field equations \eqref{fl1},
\begin{subequations}
\label{feq11}%
\begin{align}
&  3H^{2}f^{\prime}(Q)+\frac{1}{2}\left(  f(Q)-Qf^{\prime}(Q)\right)
=0\label{con1a}\\
&  -\frac{2}{N}\frac{d}{dt}\left(  f^{\prime}\left(  Q\right)  H\right)
-3H^{2}f^{\prime}(Q)-\frac{1}{2}\left(  f(Q)-Qf^{\prime}(Q)\right)  =0,
\label{con2a}%
\end{align}
where
\end{subequations}
\begin{equation}
Q=-6H^{2} \label{Qcon1}%
\end{equation}
and $H= \frac{1}{N} \frac{\dot{a}}{a}$. The Eq. \eqref{kl.01} is identically
satisfied for $\Gamma_{1}$ hence we characterize this connection as being in
the coincident gauge.

It is a known result that the field equations (\ref{con1a}), (\ref{con2a}) and
(\ref{Qcon1}) follow from the variation of the point-like Lagrangian%
\begin{equation}
L\left(  \Gamma_{1}\right)  =-\frac{6}{N}af^{\prime}(Q)\dot{a}^{2}%
+Na^{3}\left(  f(Q)-Qf^{\prime}(Q)\right)  . \label{Lag}%
\end{equation}
Indeed, when the variation is performed with respect to the lapse function
$N$, it results in the constraint equation (\ref{con1a}). Moreover, by varying
$L\left(  \Gamma_{1}\right) $ with respect to the scale factor $a$ and the
non-metricity scalar $Q$, we obtain the equations (\ref{con2a}) and
(\ref{Qcon1}). We remark that the $f\left(  Q\right)  $-theory for the first
connection is a second-order theory of gravity and equivalent with the
teleparallel $f\left(  T\right) $-cosmology for a spatially flat FLRW
\cite{Ferraro}.

At this point we can explain why we use the term ``coincident gauge''
loosely in what regards connection $\Gamma_{1}$. First of all, let us note
that the function $\gamma_{1}$ remains arbitrary, that is, it is not specified
by any equation of the motion. This is why we refer to it as non-dynamical.
What is more, the transformation which sends $\Gamma_{1}$ to the actual
coincident gauge, have the property of leaving Lagrangian \eqref{Lag}
invariant. We can set $\Gamma_{1}$ to zero, by performing the following series
of transformations on the base manifold: (a) a three-dimensional spatial
transformation from spherical coordinates $r,\phi,\theta$ to Cartesian ones
$x,y,z$ (this makes zero all components of $\Gamma_{1}$, save for the pure
temporal one) and (b) a pure time transformation that absorbs $\gamma_{1}(t)$,
in this case a mapping $t\mapsto\tau(t)$ satisfying $\ddot{\tau}/\dot{\tau
}=\gamma_{1}$. If we apply this transformation to the FLRW metric and
recalculate the minisuperspace Lagrangian, we shall obtain once more
Lagrangian \eqref{Lag}, or more precisely a Lagrangian that is connected to
\eqref{Lag} through a pure time transformation, i.e. involving a different lapse function. In particular, the FLRW metric
will take the usual form in Cartesian coordinates
\begin{equation}
ds^{2} = -\bar{N}(\tau)^{2} d\tau^{2} + a(\tau)^{2}(dx^{2} + dy^{2} + dz^{2}),
\end{equation}
with $\bar{N}(\tau)=N(t) \dot{\tau}(t)= N(t) \exp\left( \int
\!\!\gamma_{1} dt\right) $. Since cosmological Lagrangians like \eqref{Lag}
are known for their property of being parametrization invariant, i.e.
invariant under changes in time, this means that \eqref{Lag} corresponds to the same form of Lagrangian as if we had set $\Gamma_{1}=0$. The same is not true for the other two connections that we study.

\subsubsection{Connection $\Gamma_{2}$}

For the second connection, namely $\Gamma_{2}$, the additional non-zero
components are as follows%
\[
\Gamma_{\;tt}^{t}=\frac{\dot{\gamma}(t)}{\gamma(t)}+\gamma(t),\quad
\Gamma_{\;tr}^{r}=\Gamma_{\;rt}^{r}=\Gamma_{\;t\theta}^{\theta}=\Gamma
_{\;\theta t}^{\theta}=\Gamma_{\;t\varphi}^{\varphi}=\Gamma_{\;\varphi
t}^{\varphi}=\gamma(t).
\]
Thus, the gravitational field equations \eqref{fl1} can be expressed as
\begin{subequations}
\label{feq12}%
\begin{align}
&  \frac{3\dot{a}^{2}f^{\prime}(Q)}{a^{2}N^{2}}+\frac{1}{2}\left(
f(Q)-Qf^{\prime}(Q)\right)  +\frac{3\gamma\dot{Q}f^{\prime\prime}(Q)}{2N^{2}%
}=0,\label{feq12b}\\
&  -\frac{2}{N}\frac{d}{dt}\left(  \frac{f^{\prime}(Q)\dot{a}}{Na}\right)
-\frac{3\dot{a}^{2}}{N^{2}a^{2}}f^{\prime}(Q)-\frac{1}{2}\left(
f(Q)-Qf^{\prime}(Q)\right)  +\frac{3\gamma\dot{Q}f^{\prime\prime}(Q)}{2N^{2}%
}=0,
\end{align}
where
\end{subequations}
\begin{equation}
Q=-\frac{6\dot{a}^{2}}{N^{2}a^{2}}+\frac{3\gamma}{N^{2}}\left(  \frac{3\dot
{a}}{a}-\frac{\dot{N}}{N}\right)  +\frac{3\dot{\gamma}}{N^{2}}. \label{Qcon2}%
\end{equation}
Finally, the equation of motion governing the connection, \eqref{kl.01},
reads
\begin{equation}
\dot{Q}^{2}f^{\prime\prime\prime}(Q)+\left[  \ddot{Q}+\dot{Q}\left(
\frac{3\dot{a}}{a}-\frac{\dot{N}}{N}\right)  \right]  f^{\prime\prime}(Q)=0.
\label{feq22}%
\end{equation}
Notice that for $\Gamma_{2}$, Eq. \eqref{kl.01} now is not trivialized. The
connection $\Gamma_{2}$ is a dynamical one.

To formulate a point-like Lagrangian for the aforementioned set of
differential equations, we introduce the new scalar field $\psi$, which is
related to the function $\gamma$ as $\gamma\left(  t\right)  =\dot{\psi}$.
Subsequently, it becomes evident that the function%
\begin{equation}
L\left(  \Gamma_{2}\right)  =-\frac{3a\dot{a}^{2}f^{\prime}(Q)}{N}+\frac{N}%
{2}a^{3}\left(  f(Q)-Qf^{\prime}(Q)\right)  -\frac{3a^{3}\dot{\psi}\dot
{Q}f^{\prime\prime}(Q)}{2N}, \label{ds1}%
\end{equation}
is a point-like Lagrangian for the field equations (\ref{feq12b}%
)-(\ref{feq22}), if you variate with respect to $N$, $a$, $Q$ and $\psi$
respectively. The reason for introducing a field $\psi$ so that $\gamma
=\dot{\psi}$ lies in the transformation law of the function $\gamma$. If we
make a change in the time variable $t\mapsto\bar{t}$, from the transformation
law of the connection we obtain $\gamma(t) dt = \gamma(\bar{t}) d\bar{t}$.
That is, the function $\gamma$ transforms as a ``velocity'', we thus need to
define $\psi=\int\!\! \gamma(t) dt$ in order for the newly introduced field to
be a scalar under transformations in time.

We proceed with the definition of the second scalar field $\phi=f^{\prime
}\left(  Q\right) $, which results in
\begin{equation}
L\left(  \Gamma_{2}\right)  =-\frac{3}{N}\phi a\dot{a}^{2}-\frac{3}{2N}%
a^{3}\dot{\phi}\dot{\psi} +\frac{N}{2}a^{3}V\left(  \phi\right)  , \label{sd1}%
\end{equation}
where $V\left(  \phi\right)  =\left(  f(Q)-Qf^{\prime}(Q)\right)  $.

If, for Lagrangian $L\left(  \Gamma_{2}\right) $, we calculate the matrix
$G_{ij}=\frac{\partial L\left(  \Gamma_{2}\right)  }{\partial\dot{q}%
^{i}\partial\dot{q}^{j}}$, where $\mathbf{q}=(a,\phi,\psi)$, we are going to
see that it is not degenerate, i.e. $\det(G_{ij}) \neq0$. This leads us to the
conclusion that there exists no coordinate transformation capable of
eliminating the scalar fields. Consequently, it can be inferred that the
$f\left(  Q\right)  $-theory corresponding to connection $\Gamma_{2}$
propagates as many degrees of freedom as a sixth-order theory,
characterized by the presence of two scalar fields, $\phi$ and $\psi$.
We need to clarify of course that $f(Q)$ gravity is a second order
theory since this is the maximum order of derivatives appearing in the field
equations. We are simply referring to the fact that, in what regards the
degrees of freedom, this is a similar setting as one produced by a $3\times2$
order theory.

In the context of the scalar field description, and setting the lapse function
\thinspace$N=1$, the field equations can be expressed in the following form%
\begin{equation}
3\phi H^{2}+\frac{1}{2}V\left(  \phi\right)  +\frac{3}{2}\dot{\phi}\dot{\psi
}=0, \label{co.01}%
\end{equation}%
\begin{equation}
2\phi\left(  2\dot{H}+3H^{2}\right)  +4H\dot{\phi}-3\dot{\phi}\dot{\psi
}+V\left(  \phi\right)  =0, \label{co.02}%
\end{equation}%
\begin{equation}
V^{\prime}(\phi)-6H^{2}+9H\dot{\psi}+3\ddot{\psi}=0, \label{co.03}%
\end{equation}
and%
\begin{equation}
\ddot{\phi}+3H\dot{\phi}=0. \label{co.04}%
\end{equation}
The prime in the above relations denoting differentiation with respect to
$\phi$.

Equation (\ref{co.04}) represents nothing more than the definition of the
non-metricity scalar $Q$, which is indicated by expression (\ref{Qcon2}).
Additionally, equation (\ref{co.03}) is the equation of motion for the
connection, as described by equation (\ref{feq22}).

The Lagrangian \eqref{sd1} is quadratic in the velocities and its point
symmetries can be calculated in a straightforward manner. They are related to
the minisuperspace geometry $G_{ij}$ and the potential part of $L\left(
\Gamma_{2}\right) $ \cite{geomLP}. Due to the fact that Lagrangians like
$L\left(  \Gamma_{2}\right) $ have by construction a zero Hamiltonian $H=0$,
one may extend the search of integrals of motion $I$ that satisfy
$\{I,H\}=\omega(q) H$. Point symmetries behind this type of conserved
quantities are revealed as the isometries of the metric $\tilde{G}_{ij}= N
a^{3}V\left(  \phi\right)  G_{ij}$, \cite{tchris}. Apart from the obvious
symmetry $X_{1} = \partial_{\psi}$, which implies that the momentum for $\psi$
is conserved, we have two additional symmetries which emerge under specific
conditions over the potential. First we notice that the vector
\begin{equation}
X_{2} = -\frac{1}{6} a \left( \mu+1 +2 \sigma\ln\phi\right)  \partial_{a} +
\phi\, \partial_{\phi}+ \frac{4}{3} \sigma\ln a \, \partial_{\psi}%
\end{equation}
is an isometry of $\tilde{G}_{ij}$ if the potential is of the form
\begin{equation}
\label{potX2}V(\phi)=\lambda\, \phi^{\mu}e^{\sigma(\ln\phi)^{2}},
\end{equation}
where $\lambda$, $\sigma$ and $\mu$ are constants of integration. The last
symmetry is given by
\begin{equation}
X_{3} = -\frac{1}{6} a \left[ (\mu+1) \ln\phi+\sigma\right]  \partial_{a} +
\phi\ln\phi\, \partial_{\phi}+\left[ \frac{2}{3} (\mu+1) \ln a-\psi\right]
\partial_{\psi}%
\end{equation}
which exists when the potential is
\begin{equation}%
\begin{split}
V(\phi) = \lambda\, \phi^{\mu}(\ln\phi)^{\sigma}.
\end{split}
\end{equation}
It is not in the scope of this work to delve in a complete symmetry analysis
of the ensuing Lagrangians, so we shall only restrict to mention that $X_{1}$
commutes with $X_{2}$ which means that the corresponding integrals of motion,
together with the Hamiltonian, ensure the complete integrability of the system
for potentials of the form \eqref{potX2}. Finally, we notice that all three
vectors $X_{1}$, $X_{2}$ and $X_{3}$ coexist as symmetries, if $\sigma=0$ in
the above relations, i.e. when $V(\phi)= \lambda\phi^{\mu}$. It is easy to
check that, such a potential is either related to the power-law function
$f(Q)\sim Q^{\frac{\mu}{\mu-1}}$.

\subsubsection{Connection $\Gamma_{3}$}

For connection $\Gamma_{3}$, the additional non-zero components read%
\[
\Gamma_{\;tt}^{t}=-\frac{\dot{\gamma}(t)}{\gamma(t)},\quad\Gamma_{\;rr}%
^{t}=\gamma(t),\quad\Gamma_{\;\theta\theta}^{t}=\gamma(t)r^{2},\quad
\Gamma_{\;\varphi\varphi}^{t}=\gamma(t)r^{2}\sin^{2}\theta.
\]

The cosmological field equations are
\begin{subequations}
\label{feq13}%
\begin{align}
&  \frac{3\dot{a}^{2}f^{\prime}(Q)}{a^{2}N^{2}}+\frac{1}{2}\left(
f(Q)-Qf^{\prime}(Q)\right)  -\frac{3\gamma\dot{Q}f^{\prime\prime}(Q)}{2a^{2}%
}=0,\label{feq13b}\\
&  -\frac{2}{N}\frac{d}{dt}\left(  \frac{f^{\prime}(Q)\dot{a}}{Na}\right)
-\frac{3\dot{a}^{2}}{N^{2}a^{2}}f^{\prime}(Q)-\frac{1}{2}\left(
f(Q)-Qf^{\prime}(Q)\right)  +\frac{\gamma\dot{Q}f^{\prime\prime}(Q)}{2a^{2}%
}=0,
\end{align}
Moreover, the equation of motion for the connection reads%
\end{subequations}
\begin{equation}
\dot{Q}^{2}f^{\prime\prime\prime}(Q)+\left[  \ddot{Q}+\dot{Q}\left(
\frac{\dot{a}}{a}+\frac{\dot{N}}{N}+\frac{2\dot{\gamma}}{\gamma}\right)
\right]  f^{\prime\prime}(Q)=0. \label{feq23}%
\end{equation}
where now
\begin{equation}
Q=-\frac{6\dot{a}^{2}}{N^{2}a^{2}}+\frac{3\gamma}{a^{2}}\left(  \frac{\dot{a}%
}{a}+\frac{\dot{N}}{N}\right)  +\frac{3\dot{\gamma}}{a^{2}}. \label{Qcon3}%
\end{equation}

We proceed with the identical methodology as previously outlined, introducing
the new scalar field $\gamma=\frac{1}{\dot{\Psi}}$. Thus, the following
function%
\begin{equation}
L\left(  \Gamma_{3}\right)  =-\frac{3a\dot{a}^{2}f^{\prime}(Q)}{N}+\frac{N}%
{2}a^{3}\left(  f(Q)-Qf^{\prime}(Q)\right)  -\frac{3}{2}aN\frac{\dot
{Q}f^{\prime\prime}(Q)}{\dot{\Psi}}%
\end{equation}
is the point-like Lagrangian for the field equations. Once more, we define the
newly introduced field $\Psi$ so that it transforms as a scalar under time
transformations. It is easy to check that from the connection, when
$t\mapsto\bar{t}$, we get the transformation law $\frac{dt}{\gamma(t)}=
\frac{d\bar{t}}{\gamma(\bar{t})}$.

Moreover, we also introduce the second scalar field $\phi=f^{\prime}\left(
Q\right)  $, thus the latter point-like Lagrangian to become%
\begin{equation}
L\left(  \Gamma_{3}\right)  =-\frac{3}{N}\phi a\dot{a}^{2}+\frac{N}{2}%
a^{3}V\left(  \phi\right)  -\frac{3}{2}aN\frac{\dot{\phi}}{\dot{\Psi}}
\label{sd2}%
\end{equation}
where the resulting Euler-Lagrange equations, for $N=1$, are written as
follows
\begin{equation}
3\phi H^{2}+\frac{1}{2}V\left(  \phi\right)  -\frac{3\dot{\phi}}{2a^{2}
\dot{\Psi}}=0,
\end{equation}%
\begin{equation}
2\phi\left(  2\dot{H}+3H^{2}\right)  +4H\dot{\phi}- \frac{1}{a^{2}}\frac
{\dot{\phi}}{\dot{\Psi}}+V\left(  \phi\right)  =0,
\end{equation}%
\begin{equation}
\left(  6H^{2}-V^{\prime}(\phi)\right)  a^{2}\dot{\Psi}^{2} - 3 H\dot{\Psi}+
3\ddot{\Psi}=0,
\end{equation}
and%
\begin{equation}
2\dot{\phi}\ddot{\Psi}-\dot{\Psi}\left(  H\dot{\phi}+\ddot{\phi}\right)  =0.
\end{equation}

Similar to the previous approach, the field equations demonstrate equivalence
to those of a two-scalar field model. In essence, it can be deduced that the
$f(Q)$-theory corresponding to connection $\Gamma_{3}$ conforms to a
sixth-order gravitational theory characterized by the two scalar fields $\phi$
and $\Psi$.

\subsection{Non-zero spatial curvature $k\neq0$}

For the FLRW with $k\neq0$, the nonzero components of the connection
$\Gamma_{4}$ are expressed as
\begin{equation}%
\begin{split}
&  \Gamma_{\;tr}^{r}=\Gamma_{\;rt}^{r}=\Gamma_{\;t\theta}^{\theta}%
=\Gamma_{\;\theta t}^{\theta}=\Gamma_{\;t\varphi}^{\varphi}=\Gamma_{\;\varphi
t}^{\varphi}=-\frac{k}{\gamma(t)},\quad\Gamma_{\;rr}^{r}=\frac{kr}{1-kr^{2}%
},\\
&  \Gamma_{\;\theta\theta}^{r}=-r\left(  1-kr^{2}\right)  , \quad
\Gamma_{\;\varphi\varphi}^{r}=-r\sin^{2}(\theta)\left(  1-kr^{2}\right)
\quad\Gamma_{\;r\theta}^{\theta}=\Gamma_{\;\theta r}^{\theta}=\Gamma
_{\;r\varphi}^{\varphi}=\Gamma_{\;\varphi r}^{\varphi}=\frac{1}{r},\\
&  \Gamma_{\;\varphi\varphi}^{\theta}=-\sin\theta\cos\theta,\quad
\Gamma_{\;\theta\varphi}^{\varphi}=\Gamma_{\;\varphi\theta}^{\varphi}%
=\cot\theta,
\end{split}
\end{equation}

and%
\[
\Gamma_{\;tt}^{t}=-\frac{k+\dot{\gamma}(t)}{\gamma(t)},\quad\Gamma_{\;rr}%
^{t}=\frac{\gamma(t)}{1-kr^{2}}\quad\Gamma_{\;\theta\theta}^{t}=\gamma
(t)r^{2},\quad\Gamma_{\;\varphi\varphi}^{t}=\gamma(t)r^{2}\sin^{2}(\theta).
\]

\subsubsection{Connection $\Gamma_{4}$}

As a result, for this connection, the calculation of the non-metricity scalar
is produced
\begin{equation}
Q=-\frac{6\dot{a}^{2}}{N^{2}a^{2}}+\frac{3\gamma}{a^{2}}\left(  \frac{\dot{a}%
}{a}+\frac{\dot{N}}{N}\right)  +\frac{3\dot{\gamma}}{a^{2}}+k\left[  \frac
{6}{a^{2}}+\frac{3}{\gamma N^{2}}\left(  \frac{\dot{N}}{N}+\frac{\dot{\gamma}%
}{\gamma}-\frac{3\dot{a}}{a}\right)  \right]  . \label{Qconk}%
\end{equation}

The gravitational field equations are
\begin{subequations}
\label{feq1k}%
\begin{align}
&  \frac{3\dot{a}^{2}f^{\prime}(Q)}{a^{2}N^{2}}+\frac{1}{2}\left(
f(Q)-Qf^{\prime}(Q)\right)  -\frac{3\gamma\dot{Q}f^{\prime\prime}(Q)}{2a^{2}%
}+3k\left(  \frac{f^{\prime}(Q)}{a^{2}}-\frac{\dot{Q}f^{\prime\prime}%
(Q)}{2\gamma N^{2}}\right)  =0,\label{feq1kb}\\
&  -\frac{2}{N}\frac{d}{dt}\left(  \frac{f^{\prime}(Q)\dot{a}}{Na}\right)
-\frac{3\dot{a}^{2}}{N^{2}a^{2}}f^{\prime}(Q)-\frac{1}{2}\left(
f(Q)-Qf^{\prime}(Q)\right)  +\frac{\gamma\dot{Q}f^{\prime\prime}(Q)}{2a^{2}%
}\nonumber\\
&  -k\left(  \frac{f^{\prime}(Q)}{a^{2}}+\frac{3\dot{Q}f^{\prime\prime}%
(Q)}{2\gamma N^{2}}\right)  =0,
\end{align}
\end{subequations}
where the equation of motion for the connection reads
\begin{equation}%
\begin{split}
\dot{Q}^{2}f^{\prime\prime\prime}(Q)\left(  1+\frac{ka^{2}}{N^{2}\gamma^{2}%
}\right)  + &  \Bigg[ \ddot{Q}\left(  1+\frac{ka^{2}}{N^{2}\gamma^{2}}\right)
+\dot{Q}\Bigg( \left(  1+\frac{3ka^{2}}{N^{2}\gamma^{2}}\right)  \frac{\dot
{a}}{a}\\
&  +\left(  1-\frac{ka^{2}}{N^{2}\gamma^{2}}\right)  \frac{\dot{N}}{N}%
+\frac{2\dot{\gamma}}{\gamma}\Bigg) \Bigg] f^{\prime\prime}(Q)=0.\label{feq2k}%
\end{split}
\end{equation}

At this point, it is crucial to highlight the observation that when$\ k=0$,
the aforementioned field equations simplify to those of the spatially flat
FLRW model for connection $\Gamma_{3}$. Due to this resemblance, we anticipate
that $\gamma=\frac{1}{\dot{\Psi}}$. As anticipated, with this alteration of
variables, the subsequent function unfolds the point-like Lagrangian%
\begin{equation}%
\begin{split}
L\left(  \Gamma_{4}\right)  =  & \frac{3k}{2N}a^{3}f^{\prime\prime}%
(Q)\dot{\Psi}\dot{Q} -\frac{3a\dot{a}^{2}f^{\prime}(Q)}{N} +\frac{N}{2}%
a^{3}\left(  f(Q)-Qf^{\prime}(Q)\right)  +3kN a f^{\prime}(Q)\\
&  -\frac{3}{2}aN\frac{\dot{Q}f^{\prime\prime}(Q)}{\dot{\Psi}}.\label{sd4}%
\end{split}
\end{equation}
Hence, for $\phi=f^{\prime}\left(  Q\right)  $ and $N\left(  t\right)  =1,$
the gravitational field equations read%
\begin{equation}
3\phi\left(  H^{2}+\frac{k}{a^{2}}\right)  +\frac{1}{2}V\left(  \phi\right)
-\frac{3\dot{\phi}}{2a^{2}\dot{\Psi}}-\frac{3}{2}k \dot{\phi}\dot{\Psi}=0,
\end{equation}%
\begin{equation}
2 \phi\left(  2\dot{H}+ 3H^{2}+\frac{k}{a^{2}}\right) + V(\phi) +4H\dot{\phi}
+ 3k\dot{\phi}\dot{\Psi} - \frac{\dot{\phi}}{a^{2}\dot{\Psi}}=0,
\end{equation}%
\begin{equation}
\left(  6H^{2}-V^{\prime}(\phi)+3k\left(  \ddot{\Psi}+3H\dot{\Psi}\right)
\right)  \frac{a^{2}\dot{\Psi}^{2}}{3} - H\dot{\Psi}+\ddot{\Psi}-2k\dot{\Psi
}^{2}=0,
\end{equation}
and%
\begin{equation}
H\dot{\phi}\dot{\Psi}+\dot{\Psi}\ddot{\Phi}+a^{2}k\dot{\Psi}^{3}\left(
3H\dot{\phi}+\ddot{\phi}\right)  -2\dot{\phi}\ddot{\Psi}=0.
\end{equation}

\section{Analytic solution for $\Gamma_{2}$}

\label{sec4}

We shall proceed to construct the general solution for the field equations
corresponding to connection $\Gamma_{2}$. We select the lapse function
$N\left(  t\right)  =a^{3}$, where the field equations are written in the
following form%
\begin{equation}
-\frac{1}{2}a^{3} V(\phi)-\frac{3\phi}{a^{5}}\dot{a}^{2}-\frac{3}{2a^{3}}%
\dot{\phi}\dot{\psi}=0,
\end{equation}%
\begin{equation}
-\frac{3}{2}a^{5}V\left(  \phi\right)  +\frac{15}{a^{3}}\phi\dot{a}^{2}%
-\frac{6}{a^{2}}\dot{a}\dot{\phi}+\frac{9}{2a}\dot{\phi}\dot{\psi}-\frac
{6}{a^{2}}\phi\ddot{a}=0,
\end{equation}%
\begin{equation}
3\left(  \frac{\dot{a}}{a}\right)  ^{2}-\frac{1}{2}a^{6}V_{,\phi}-\frac{3}%
{2}\ddot{\psi}=0,
\end{equation}
and%
\begin{equation}
\ddot{\phi}=0\text{.} \label{dd3}%
\end{equation}

Easily we calculate $\phi\left(  t\right)  =\phi_{1}\left(  t-t_{0}\right)  ,$
leading to the emergence of a second-order differential equation for the scale
factor
\begin{equation}
\ddot{\beta}+\frac{\dot{\beta}}{\left(  t-t_{0}\right)  }+\frac{1}{2}%
e^{6\beta}\frac{V\left(  \phi_{1}\left(  t-t_{0}\right)  \right)  }{\phi
_{1}\left(  t-t_{0}\right)  }=0~\text{\ where }\beta\left(  t\right)  =\ln
a\left(  t\right)  . \label{dd2}%
\end{equation}

We proceed with the second change of variable $t-t_{0}=e^{T}$. We arrive at
the resulting second-order differential equation%
\begin{equation}
\frac{d^{2}\beta}{dT^{2}}+\frac{1}{2\phi_{1}}e^{6\beta}e^{T}V\left(  T\right)
=0. \label{dd1}%
\end{equation}

In order to demonstrate the construction of an exact solution, let us consider
the power-law theory $f\left(  Q\right)  =f_{0}Q^{\frac{n-1}{n-2}}$. This
allows us to deduce the power-law scalar field potential $V\left(
\phi\right)  =V_{0}\phi^{n-2}$.

Consequently, equation, (\ref{dd1}) reads
\begin{equation}
\ddot{\beta}+\frac{V_{0}}{2\phi_{1}}e^{6\beta}e^{nT}=0
\end{equation}
with analytic solution%
\begin{equation}
\beta\left(  T\right)  =\frac{1}{6}\left[  -nT+\ln\left[  \frac{\left(
n^{2}+\beta_{1}\right) \phi_{1}^{2-n}}{6V_{0}}\left(  \cosh\left(  \frac
{\sqrt{n^{2}+\beta_{1}}}{2}\left(  T-T_{0}\right)  \right)  \right) ^{-2}
\right]  \right]  .
\end{equation}

\section{Anisotropic cosmology}

\label{sec5}

We extend our analysis by considering anisotropic cosmological models.
Specifically, we assume that line element \cite{dm3}
\begin{equation}
ds^{2}=-N^{2} dt^{2}+a^{2}\left(  t\right)  \left(  e^{2b\left(  t\right)
}dx^{2}+e^{-b\left(  t\right)  }\left(  dy^{2}+S^{2}\left(  y\right)
dz^{2}\right)  \right)  , \label{ks.01}%
\end{equation}
where we have either $S(y)=\sin y$ or $S(y)=\sinh y$ with $k=-S^{\prime\prime
}(y)/S(y)$. The $a\left(  t\right)  $ is the scale factor describing the size
of the universe and $b\left(  t\right)  $ is the anisotropic parameter. For
$k=+1$, the line element (\ref{ks.01}) describes a Kantowski-Sachs geometry,
while $k=-1$ corresponds to the locally rotational (LRS) Bianchi III
spacetime. On the other hand, for $k=0$, geometry (\ref{ks.01}) takes the form
of LRS Bianchi I spacetime. Self-similar anisotropic solutions for the latter
geometry in $f\left(  Q\right)  $-gravity investigated in \cite{dm3}. It was
found that there exist two-families of flat connections which inherits the
symmetries of the background geometry.

Indeed, spacetime (\ref{ks.01}) admits a four-dimensional Lie algebra
consisted by the vector fields%

\[
\xi_{1}=\partial_{z}~,~\xi_{2}=\cos z\, \partial_{y}-\frac{S^{\prime}%
(y)}{S(y)}\sin z \, \partial_{z}%
\]

\[
\xi_{3}=\sin z\, \partial_{y}+\frac{S^{\prime}(y)}{S(y)} \cos z \,\partial
_{z}\text{ and }\xi_{4}=\partial_{x}.
\]

The two families of connections are as follow. Connection $\Gamma_{A}$ with
non-zero coefficients%

\begin{equation}%
\begin{split}
&  \Gamma_{\;tt}^{t}=\gamma_{2},\quad\Gamma_{\;tt}^{x}=\frac{1}{c_{1}}\left(
\dot{\gamma}_{1}-\gamma_{1}\gamma_{2}+\gamma_{1}^{2}\right)  ,\quad
\Gamma_{\;tx}^{x}=\Gamma_{\;ty}^{y}=\Gamma_{\;tz}^{z}=\gamma_{1},\quad
\Gamma_{\;yy}^{x}=-\frac{1}{c_{1}},\\
&  \Gamma_{\;xx}^{x}=\Gamma_{\;xy}^{y}=\Gamma_{\;xz}^{z}=c_{1},\quad
\Gamma_{\;zz}^{x}=-\frac{k S(y)^{2}}{c_{1}},\quad\Gamma_{\;zz}^{y}%
=-S(y)S^{\prime}(y),\quad\Gamma_{\;yz }^{z}=\frac{S^{\prime}(y)}{S(y)}.
\end{split}
\end{equation}
and connection $\Gamma_{B}$ with non-zero coefficients%

\begin{equation}%
\begin{split}
&  \Gamma_{\;tt}^{t}=-\frac{1}{\gamma_{2}}\left[  \dot{\gamma}_{2}+c_{1}%
\gamma_{1}\left(  2-c_{2}\gamma_{1}\right)  +k\right]  ,\quad\Gamma_{\;tx}%
^{t}=c_{1}\left(  1-c_{2}\gamma_{1}\right)  ,\quad\Gamma_{\;xx}^{t}=c_{1}%
c_{2}\gamma_{2},\\
&  \Gamma_{\;yy}^{t}=\gamma_{2}, \quad\Gamma_{\;zz}^{t}=\gamma_{2}%
S(y)^{2},\quad\Gamma_{\;tt}^{x}=\frac{1}{\gamma_{2}^{2}}\left[  \gamma
_{1}\left(  k+c_{1}\gamma_{1}\right)  \left(  c_{2}\gamma_{1}-1\right)
-\gamma_{2}\dot{\gamma}_{1}\right] \\
&  \Gamma_{\;tx}^{x}=-\frac{c_{2}\gamma_{1}}{\gamma_{2}}\left( k +c_{1}%
\gamma_{1}\right)  , \quad\Gamma_{\;xx}^{x}=c_{1}+c_{2} k+c_{1}c_{2}\gamma
_{1},\quad\Gamma_{\;yy}^{x}=\gamma_{1}, \quad\Gamma_{\;zz}^{x}=\gamma
_{1}S(y)^{2},\\
&  \Gamma_{\;ty}^{y}=\Gamma_{\;tz}^{z}=-\frac{k+c_{1}\gamma_{1}}{\gamma_{2}},
\quad\Gamma_{\;xy}^{y}=\Gamma_{\;xz}^{z}=c_{1},\quad\Gamma_{\;zz}%
^{y}=-S(y)S^{\prime}(y),\quad\Gamma_{\;yz }^{z}=\frac{S^{\prime}(y)}{S(y)}.
\end{split}
\end{equation}

We proceed with the presentation of the field equations and the derivation of
the minisuperspace Lagrangian.

\subsection{Connection $\Gamma_{A}$}

For the first connection, namely $\Gamma_{A}$, the non-metricity scalar is
calculated%
\begin{equation}
Q=-6H^{2}+2\frac{k}{a^{2}}e^{b}+\frac{3}{2}\frac{\dot{b}^{2}}{N^{2}}+ 9H
\frac{\gamma_{1}}{N} - 3 \frac{\gamma_{1} \dot{N}}{N^{3}} +3 \frac{\dot
{\gamma}_{1}}{N^{2}},
\end{equation}
where $H=\frac{\dot{a}}{N a}$. The corresponding field equations have the
diagonal components

$tt:$%
\begin{equation}
f^{\prime}(Q)\left(  3H^{2}+\frac{k}{a^{2}}e^{b}-\frac{3}{4}\frac{\dot{b}^{2}%
}{N^{2}}\right)  +\frac{1}{2}\left(  f(Q)-Qf^{\prime}(Q)\right)  +\frac{3}%
{2}\frac{\gamma_{1}\dot{Q}f^{\prime\prime}(Q)}{N^{2}}=0,
\end{equation}

$xx:$%
\begin{equation}%
\begin{split}
&  f^{\prime}(Q)\left(  \frac{\ddot{b}}{N^{2}}+3H\frac{\dot{b}}{N}- \frac
{\dot{b}\dot{N}}{N^{3}}-2\frac{\dot{H}}{N}-3H^{2}-\frac{k}{a^{2}}e^{b}%
-\frac{3}{4}\frac{\dot{b}^{2}}{N^{2}} \right)  -\frac{1}{2}\left(
f(Q)-Qf^{\prime}(Q)\right) \\
&  +\frac{\dot{Q}}{N}f^{\prime\prime}(Q)\left(  \frac{\dot{b}}{N} -2H+\frac
{3}{2}\frac{\gamma_{1}}{N}\right)  =0,
\end{split}
\end{equation}

$yy,zz:$%
\begin{equation}%
\begin{split}
&  f^{\prime}(Q)\left(  \frac{\ddot{b}}{N^{2}}+3H\frac{\dot{b}}{N}+4\frac
{\dot{H}}{N}+6 H^{2}+\frac{3}{2}\frac{\dot{b}^{2}}{N^{2}}\right)  + \left(
f(Q)-Qf^{\prime}(Q)\right) \\
&  + \frac{\dot{Q}}{N}f^{\prime\prime}(Q)\left(  4H+\frac{\dot{b}}{N}%
-3\frac{\gamma_{1}}{N}\right)  =0.
\end{split}
\end{equation}
Furthermore, there exist the diagonal component%
\begin{equation}
\frac{3}{2}c_{1}\dot{Q}f^{\prime\prime}=0.
\end{equation}
Finally, from the equation of motion for the connection (\ref{kl.01}) we
calculate%
\begin{equation}
f^{\prime\prime\prime}(Q) \frac{Q^{\prime2}}{N^{2}}+f^{\prime\prime}(Q)\left(
\frac{\ddot{Q}}{N^{2}}+3H\frac{\dot{Q}}{N}- \frac{\dot{Q}\dot{N}}{N^{3}%
}\right)  =0.
\end{equation}

We remark that for in the case of the vacuum, as we consider here, the
non-diagonal component $c_{1}\dot{Q}f^{\prime\prime}(Q)=0$, implies that
either $f\left(  Q\right)  $ is linear, and the theory reduced to the STGR, or
$\dot{Q}=0$, which again leads to dynamics equivalent of that of GR with a
cosmological constant. Notice that $c_{1}$ cannot be zero due to its
appearance in denominators in the components of the connection $\Gamma_{A}$.
Thus, connection $\Gamma_{A}$ can not reproduce the field equations for the
vacuum or for a perfect fluid.

\subsection{Connection $\Gamma_{B}$}

For connection $\Gamma_{B}$, for the non-metricity term we consider the case
\begin{equation}
\label{gamma1choice}\gamma_{1} = -\frac{1}{c_{2}}-\frac{k}{c_{1}}.
\end{equation}
For such a selection for the function $\gamma_{1}$ the field equations become diagonal.

The non-metricity scalar is then expressed as
\begin{align}
Q  &  =-6H^{2}+2\frac{k}{a^{2}}e^{b}+\frac{3}{2}\frac{\dot{b}^{2}}{N^{2}%
}-\frac{6kH}{\gamma_{2} N}+\frac{e^{b}}{a^{2}}\left( 2+c_{1} c_{2} e^{-3
b}\right) \left(  H N \gamma_{2} +\dot{\gamma}_{2} + \frac{\gamma_{2} \dot{N}%
}{N}\right) \nonumber\\
&  + 2 \frac{e^{b}}{a^{2}} \left(  1- c_{1} c_{2} e^{-3b} \right)  \gamma_{2}
\dot{b} + \frac{3 (2 c_{1}+ c_{2} k)^{2}}{4 c_{1} c_{2} }\frac{H}{N \gamma
_{2}} - \frac{(c_{2} k-2 c_{1})^{2}}{4 c_{1} c_{2} N^{2} \gamma_{2}} \left(
\frac{\dot{N}}{N} + \frac{\dot{\gamma_{2}}}{\gamma_{2}} \right)  \label{qq.01}%
\end{align}
The field equations are

$tt:$%
\begin{equation}%
\begin{split}
&  f^{\prime}(Q)\left(  3H^{2}+\frac{k}{a^{2}}e^{b}-\frac{3}{4}\frac{\dot
{b}^{2}}{N^{2}}\right)  +\frac{1}{2}\left(  f(Q)-Qf^{\prime}(Q)\right) \\
&  -\frac{\dot{Q}}{N}f^{\prime\prime}(Q) \left[  \frac{(c_{2} k-2 c_{1})^{2}%
}{8 c_{1} c_{2} N \gamma_{2}} - \frac{ e^{b} N \gamma_{2}}{a^{2}}\left(
1+\frac{c_{1} c_{2}}{2} e^{-3 b}\right)  \right]  =0\label{qq.02}%
\end{split}
\end{equation}

$xx:$%
\begin{equation}%
\begin{split}
&  f^{\prime}(Q) \left(  \frac{\ddot{b}}{N^{2}} +3H\frac{\dot{b}}{N} -
\frac{\dot{b}\dot{N}}{N^{3}} -2\frac{\dot{H}}{N}-3H^{2}-\frac{k e^{b}}{a^{2}%
}-\frac{3}{4}\frac{\dot{b}^{2}}{N^{2}} \right)  -\frac{1}{2}\left(
f(Q)-Qf^{\prime}(Q)\right) \\
&  + \frac{\dot{Q}}{N}f^{\prime\prime}(Q) \left[  \frac{(c_{2} k-2 c_{1})^{2}%
}{8 c_{1} c_{2} N \gamma_{2}} + \frac{ e^{b} N \gamma_{2}}{a^{2}}\left( 1 -
\frac{c_{1} c_{2}}{2} e^{-3 b}\right)  + \frac{\dot{b}}{N} -2 H \right]
=0\label{qq.03}%
\end{split}
\end{equation}

$yy,zz:$%
\begin{equation}%
\begin{split}
&  f^{\prime}(Q) \left(  \frac{\ddot{b}}{N^{2}} +3H\frac{\dot{b}}{N} -
\frac{\dot{b}\dot{N}}{N^{3}} + 4 \frac{\dot{H}}{N} + 6 H^{2}+\frac{3}{2}%
\frac{\dot{b}^{2}}{N^{2}} \right)  +\left(  f(Q)-Qf^{\prime}(Q)\right) \\
&  + \frac{\dot{Q}}{N}f^{\prime\prime}(Q) \left[  4H+ \frac{\dot{b}}{N}
\frac{(c_{2} k-2 c_{1})^{2}}{4 c_{1} c_{2} N \gamma_{2}} - \frac{c_{1} c_{2}
e^{-2 b}}{a^{2}}N\gamma_{2} \right]  =0\label{qq.03}%
\end{split}
\end{equation}
In general the equations also include a nondiagonal component $tx$,
\begin{equation}
(2 c_{1}+c_{2} k+2 c_{1} c_{2} \gamma_{1} ) \dot{Q} f^{\prime\prime}(Q) =0,
\end{equation}
which however is trivialized due to taking \eqref{gamma1choice}. The other
alternative choice, to avoid $\dot{Q}=0$ or $f^{\prime\prime}(Q)=0$, would be
to take $c_{1}=c_{2}=\gamma_{1}=0$. At a first sight this seems to be a
different case from the choice \eqref{gamma1choice} since $c_{1}$ and $c_{2}$
appear in denominators. However, if we make with the following order the
substitutions \eqref{gamma1choice} and then $c_{1}=-\frac{c_{2} k}{2}$, which
makes $\gamma_{1}=0$, then the remaining $c_{2}$ is simplified from all
denominators and we can freely set it also equal to zero. Thus, obtaining the
results of the $c_{1}=c_{2}=\gamma_{1}=0$ case.

Now, necause of \eqref{gamma1choice}, we are also left with a single equation
of motion for the connection, which we write as
\begin{equation}%
\begin{split}
&  E(1,1,1) \left( \frac{\dot{Q}^{2}}{N^{2}}f^{\prime\prime\prime}(Q) +
\frac{\ddot{Q}}{N^{2}}f^{\prime\prime}(Q) \right)  +\frac{\dot{Q}}{N}
f^{\prime\prime}(Q) \Bigg[ E(-1,1,1)\frac{\dot{N}}{N^{2}} + E(3,1,1) H\\
&  + \frac{\dot{b}}{N} \Big(E(\sigma,\mu,\nu) + E(-\sigma,-2-\mu,1-\nu)\Big) +
\frac{\dot{\gamma_{2}}}{\gamma_{2} N} \Big(E(\sigma,1,1)+E(-\sigma,1,1)
\Big) \Bigg] =0,\label{qq.05}%
\end{split}
\end{equation}
where in order to simplify the expression we introduced the following function
of time, which we choose to characterize with three parameters $\sigma$, $\mu
$, $\nu$
\begin{equation}
E(\sigma,\mu,\nu) = a \left[ 4 c_{1} c_{2} N^{2} \gamma_{2}^{2} \left( 2\,
\nu\, e^{3 b}+\mu\, c_{1} c_{2} \right) -\sigma\, a^{2} e^{2 b} (c_{2} k-2
c_{1})^{2}\right] . \label{qq.06}%
\end{equation}
The parameters $\sigma,\mu,\nu$ themselves are unimportant, i.e. they do not
appear in the end expression of \eqref{qq.05}.

We observe that in the isotropic limit, that is, $b\rightarrow0$, the field
equations of the FLRW spacetime for the connection $\Gamma_{4}$, $\left(
k\neq0\right)  $, or $\Gamma_{3}$~$\left(  k=0\right)  $, are recovered.

In a similar way as before we consider the new scalar field $\gamma_{2}%
=\frac{1}{\dot{\Psi}}$, then the following function%
\begin{equation}%
\begin{split}
L\left(  \Gamma_{B}\right)  =  &  \frac{1}{N} \left(  \frac{3}{2} a^{3}
f^{\prime}(Q) \dot{b}^{2} - 6 a f^{\prime}(Q) \dot{a}^{2} - \frac{a^{3} (c_{2}
k-2 c_{1})^{2} f^{\prime\prime}(Q)}{4 c_{1} c_{2}}\dot{Q} \dot{\Psi} \right)
\\
&  -N \left(  a e^{-2 b} \left( 2 e^{3 b}+c_{1} c_{2}\right)  f^{\prime\prime
}(Q) \frac{\dot{Q}}{\dot{\Psi}} - a^{3} \left(  f(Q)-Q f^{\prime}(Q)\right)
-2 k a e^{b} f^{\prime}(Q) \right) ,\label{qq.07}%
\end{split}
\end{equation}
is a Lagrangian function for the field equations and nonmetricity scalar
(\ref{qq.01})-(\ref{qq.06}).

Last but not least, we can define the second scalar field $\phi=f^{\prime
}\left(  Q\right)  $, such that, the cosmological Lagrangian to expressed as
\begin{equation}%
\begin{split}
L\left(  \Gamma_{B}\right)  =  &  \frac{1}{N} \left(  \frac{3}{2} a^{3}
\phi\dot{b}^{2} - 6 a \phi\dot{a}^{2} - \frac{a^{3} (c_{2} k-2 c_{1})^{2}}{4
c_{1} c_{2}}\dot{\phi} \dot{\Psi} \right) \\
&  - N \left(  a e^{-2 b} \left( 2 e^{3 b}+c_{1} c_{2}\right)  \frac{\dot
{\phi}}{\dot{\Psi}} - a^{3} V(\phi) - 2 k a e^{b} \phi\right) ,
\end{split}
\end{equation}
with $V\left(  \phi\right)  =\left(  f(Q)-Qf^{\prime}(Q)\right)  $.

We conclude that $f\left(  Q\right)  $-theory propagates as many
degrees of freedom as a sixth-order theory of gravity, two of which
are described by scalar fields. It is important to mention that there is a
coupling between the non-coincident gauge, that is, function $\gamma_{2}$, and
the anisotropic parameter $b$. Last but not least, there coincident gauge does
not provide a solution for the field equations in the limit of vacuum.

\section{Conclusions}

\label{conc}

We demonstrated the existence of minisuperspace for the $f\left(  Q\right)
$-cosmology in homogeneous spacetimes. Specifically, for the FLRW spacetime,
we established that a minisuperspace description can be established for both
the coincident gauge and the non-coincident gauge. In the coincident gauge,
the resultant field equations describe a second-order theory similar to
General Relativity, while in the non-coincident gauge, even though we
still have second order equations, there are now additional degrees of freedom
in a setting that is reminiscent of a sixth-order theory of gravity
which is reducible to a two-scalar field model plus gravity. Of
particular interest is the role of boundary terms introduced by the various
connections in the nonmetricity scalar $Q$, which introduce higher-order
derivatives in the field equations for nonlinear functions $f(Q)$.

We have explored scenarios with non-zero spatial curvature, where the
non-linear $f(Q)$-theory consistently presents similarities to a
sixth-order theory of gravity. This outcome remains consistent for the
anisotropic spacetimes in our analysis, namely the Kantowski-Sachs and the
Bianchi III spacetimes.

To underline the novelty of the minisuperspace description, we have
successfully reconstructed the analytical solution for the power-law $f(Q) =
Q^{\alpha}$ theory within a connection defined in the non-coincident gauge.

The existence of the minisuperspace description constitutes a significant
characteristic of this theory. It enables the application of Noether symmetry
analysis for deriving new analytical solutions. Furthermore, it allows us to
formulate the Wheeler-DeWitt equation and explore quantum cosmology within the
context of the non-coincident gauge.

\textbf{Data Availability Statements:} Data sharing is not applicable to this
article as no datasets were generated or analyzed during the current study.

\begin{acknowledgments}
The author thanks the support of Vicerrector\'{\i}a de Investigaci\'{o}n y
Desarrollo Tecnol\'{o}gico (Vridt) at Universidad Cat\'{o}lica del Norte
through N\'{u}cleo de Investigaci\'{o}n Geometr\'{\i}a Diferencial y
Aplicaciones, Resoluci\'{o}n Vridt No - 098/2022. AP acknowledges the
hospitality of the Ionian University while part of this work carried out.
\end{acknowledgments}

\appendix

\section{Lagrange multiplier}

\label{appen}

In this Appendix we show the process by which point-like Lagrangians can be
derived through the incorporation of a Lagrange multiplier within the Action Integral.

Let us assume the Action Integral%
\[
S_{f\left(  Q\right)  }=\int d^{4}x\sqrt{-g}\left(  f\left(  Q\right)
+\lambda\left(  Q-\hat{Q}\right)  \right)
\]
where $\lambda$ is a Lagrange multiplier and $Q\equiv\hat{Q}$. Variation with
respect to $Q$ gives $\lambda=-f^{\prime}\left(  Q\right)  $, from where we
end with the Action%
\begin{equation}
S_{f\left(  Q\right)  }=\int d^{4}x\sqrt{-g}\left[  \left(  f\left(  Q\right)
-f^{\prime}\left(  Q\right)  Q\right)  +f^{\prime}\left(  Q\right)  \hat
{Q}\right]  . \label{mm.01}%
\end{equation}

\subsection{Connection $\Gamma_{1}$}

For connection $\Gamma_{1}$, and $Q=-6H^{2}$, expression (\ref{mm.01}) reads
\[
S_{f\left(  Q\right)  }=\int dt\left[  Na^{3}\left(  f\left(  Q\right)
-f^{\prime}\left(  Q\right)  Q\right)  -\frac{6}{N}a\dot{a}^{2}f^{\prime
}\left(  Q\right)  \right]  .
\]

We conclude that the gravitational point-like Lagrangian function is%
\[
L\left(  \Gamma_{1}\right)  =-\frac{6}{N}a\dot{a}^{2}f^{\prime}\left(
Q\right)  +Na^{3}\left(  f\left(  Q\right)  -f^{\prime}\left(  Q\right)
Q\right)  .
\]

\subsection{Connection $\Gamma_{2}$}

For connection $\Gamma_{2}$ and the definition of the non-metricity scalar
expression (\ref{mm.01}) is written as
\begin{align*}
S_{f\left(  Q\right)  }  &  =\int dt\left[  Na^{3}\left(  f\left(  Q\right)
-f^{\prime}\left(  Q\right)  Q\right)  -\frac{6}{N}a\dot{a}^{2}f^{\prime
}\left(  Q\right)  \right] \\
&  +\int dt\left[  3\gamma f^{\prime}\left(  Q\right)  \left(  \frac{3}%
{N}a^{2}\dot{a}-a^{3}\frac{\dot{N}}{N^{2}}\right)  \right]  +\int dt\left[
\frac{3}{N}f^{\prime}\left(  Q\right)  a^{3}\dot{\gamma}\right]
\end{align*}
Integration by parts for the last term gives
\[%
\begin{split}
\int dt\left[  \frac{3}{N}f^{\prime}\left(  Q\right)  a^{3}\dot{\gamma
}\right]  =  &  -\int dt\left[  -\frac{3\dot{N}}{N^{2}}f^{\prime}\left(
Q\right)  a^{3}\gamma+\frac{9}{N}f^{\prime}\left(  Q\right)  a^{2}\dot
{a}\gamma+\frac{3}{N}f^{\prime\prime}\left(  Q\right)  a^{3}\dot{Q}%
\gamma\right] \\
&  + \text{surface terms} .
\end{split}
\]

Therefore the point-like Lagrangian of the field equations is%
\[
L\left(  \Gamma_{2}\right)  =-\frac{3a\dot{a}^{2}f^{\prime}(Q)}{N}+\frac{N}%
{2}a^{3}\left(  f(Q)-Qf^{\prime}(Q)\right)  -\frac{3a^{3}\dot{\psi}\dot
{Q}f^{\prime\prime}(Q)}{2N} ,.
\]
with the use of $\gamma=\dot{\psi}$.

\subsection{Connection $\Gamma_{3}$}

For the third connection we apply the same procedure as before%
\begin{align*}
S_{f\left(  Q\right)  }  &  =\int dt\left[  Na^{3}\left(  f\left(  Q\right)
-f^{\prime}\left(  Q\right)  Q\right)  -\frac{6}{N}a\dot{a}^{2}f^{\prime
}\left(  Q\right)  \right] \\
&  +\int dt\left[  3a\gamma f^{\prime}\left(  Q\right)  \left(  \frac{\dot{a}%
}{a}N+\dot{N}\right)  \right]  +\int dt\left[  3f^{\prime}\left(  Q\right)
aN\dot{\gamma}\right]
\end{align*}
where now%
\[%
\begin{split}
\int dt\left[  3f^{\prime}\left(  Q\right)  aN\dot{\gamma}\right]  =  & -\int
dt\left[  3f^{\prime}\left(  Q\right)  a\dot{N}\gamma+3\dot{a}N\gamma
f^{\prime}\left(  Q\right)  +3a\gamma\dot{Q}f^{\prime\prime}\left(  Q\right)
\right] \\
&  + \text{surface terms}.
\end{split}
\]

We conclude that the gravitational point-like Lagrangian is
\[
L\left(  \Gamma_{3}\right)  =-\frac{3a\dot{a}^{2}f^{\prime}(Q)}{N}+\frac{N}%
{2}a^{3}\left(  f(Q)-Qf^{\prime}(Q)\right)  -\frac{3Na\dot{Q}f^{\prime\prime
}(Q)}{2\dot{\Psi}},
\]
with $\gamma=\frac{1}{\dot{\Psi}}$.

\subsection{Connection $\Gamma_{4}$}

For the fourth connection $\Gamma_{4}$, with $k\neq0$, expression
(\ref{mm.01}) reads%
\begin{align*}
S_{f\left(  Q\right)  }  &  =\int dt\left[  Na^{3}\left(  f\left(  Q\right)
-f^{\prime}\left(  Q\right)  Q\right)  -6 a f^{\prime}(Q)\left(  \frac{\dot
{a}^{2}}{N} -k N\right)  \right] \\
&  +\int dt\left[  3a\gamma f^{\prime}\left(  Q\right)  \left(  \frac{\dot{a}%
}{a}N+\dot{N}\right)  \right]  +\int dt\left[  3f^{\prime}\left(  Q\right)
aN\dot{\gamma}\right] \\
&  +\int dt\left[  \frac{3k a^{3}f^{\prime}(Q) }{\gamma N}\left(  \frac
{\dot{N}}{N}+\frac{\dot{\gamma}}{\gamma}-\frac{3\dot{a}}{a}\right)  \right]  .
\end{align*}
Consequently, integration by parts provides the Lagrangian function
\[%
\begin{split}
L\left(  \Gamma_{4}\right)  =  &  -\frac{3a\dot{a}^{2}f^{\prime}(Q)}{N}%
+\frac{N}{2}a^{3}\left(  f(Q)-Qf^{\prime}(Q)\right)  -\frac{3Na\dot
{Q}f^{\prime\prime}(Q)}{2\dot{\Psi}}+3kN a f^{\prime}(Q)\\
&  +\frac{3k}{2N}a^{3}\dot{\Psi}\dot{Q}f^{\prime\prime}(Q),
\end{split}
\]
where once more $\gamma=\frac{1}{\dot{\Psi}}$.

\end{document}